\documentclass[twocolumn,aps,floatfix]{revtex4}

\usepackage{amssymb} \usepackage{graphicx} \usepackage{amsmath}
\usepackage[T1]{fontenc} \usepackage{pstricks} \usepackage{subfigure}
\usepackage[colorlinks=true,citecolor=blue,urlcolor=red]{hyperref}
\usepackage[normalem]{ulem}
\usepackage{hyperref}
\usepackage{doi}
\usepackage{comment}

\begin{document}

\title{Dynamic formation of supersolid phase in a mixture of ultracold bosonic and fermionic atoms}

\author{Maciej Lewkowicz,$\,^1$ Tomasz Karpiuk,$\,^2$  Mariusz Gajda,$\,^3$  and Miros{\l}aw Brewczyk$\,^2$ }

\affiliation{
\mbox{$^1$ Doctoral School of Exact and Natural Sciences, University of Bia{\l}ystok, ul. K. Cio{\l}kowskiego 1K, 15-245 Białystok, Poland}
\mbox{$^2$ Wydzia{\l} Fizyki, Uniwersytet w Bia{\l}ymstoku,  ul. K. Cio{\l}kowskiego 1L, 15-245 Bia{\l}ystok, Poland}
\mbox{$^3$ Institute of Physics, Polish Academy of Sciences, Aleja Lotnik{\'o}w 32/46, PL-02668 Warsaw, Poland}
}

\begin{abstract}
We numerically study the dynamical properties of a mixture consisting of a dipolar condensate and a degenerate Fermi gas in a quasi-one-dimensional geometry. In particular, we focus on the system's response to a temporal variation in the interaction strength between bosons and fermions. When the interspecies attraction becomes sufficiently strong, we observe a phase transition to a supersolid state. This conclusion is supported by the emergence of an out-of-phase Goldstone mode in the excitation spectrum.

\end{abstract}
\maketitle

\section{Introduction}
Ultracold quantum gases provide a highly controllable platform for exploring complex many-body phenomena in quantum systems. Among the most notable recent developments is experimental  realization of quantum droplets, which arise in gases with strong dipole-dipole interactions \cite{Pfau16a,Ferlaino16} and in two-component Bose-Einstein condensates \cite{Cabrera17,Fattori18}. In particular, experiments of T. Pfau group demonstrated that, under appropriate tuning of the interparticle interactions, ensembles of droplets can form in strongly dipolar dysprosium gases. When these droplets overlap, they maintain global phase coherence \cite{Pfau16}, signaling the emergence of a supersolid—an exotic quantum phase whose existence has been discussed on theoretical grounds since the late fifties of the last century \cite{Boninsegni12,Penrose56,Gross56,Andreev69,Leggett70}.

The supersolid phase uniquely combines properties of superfluidity and crystalline structure, exhibiting simultaneous off-diagonal long-range order (phase coherence) and periodic density modulations. In dipolar Bose gases, the formation of such a phase is closely linked to the presence of a roton instability in the excitation spectrum \cite{Santos03}, which acts as a precursor to spatial ordering and droplet formation. Consequently, the coexistence of spatial structure and superfluidity has been demonstrated in arrays of quantum droplets composed of dysprosium or erbium atoms \cite{Tanzi19b,Guo19,Tanzi19a,Natale19,Bottcher19,Chomaz19,Ferlaino21,Bottcher21,Chomaz23}. Other methods of forming supersolids exploit atoms resonantly coupled to an optical cavity \cite{Leonard17,Leonard17a} or spin-orbit coupling \cite{Li17}. Supersolid systems offer a promising platform for investigating the interplay between quantum coherence, collective excitations, and spontaneous symmetry breaking in low-temperature many-body physics.

In Ref. \cite{Lewkowicz25}, we discuss the emergence of a supersolid phase in a quasi-one-dimensional mixture of a dipolar Bose-Einstein condensate and a degenerate Fermi gas. Bose-Fermi mixtures are currently the subject of intensive experimental investigation \cite{Patel23,Lippi25,Cai25}. In \cite{Lewkowicz25}, we observed a transition from a Bose-Einstein condensate to a supersolid phase induced by the presence of fermions. For this transition to occur, it is essential that bosonic and fermionic atoms attract each other. When this attraction becomes sufficiently strong, a roton instability develops, leading to modulations in both bosonic and fermionic densities. This happens because, as the boson-fermion interaction strength increases, the short-range part of the effective bosonic interaction becomes attractive, while the long-range part remains repulsive. For a fixed number of fermions, the number of resulting density peaks depends primarily on the number of bosonic atoms and can range from two up to the total number of fermions. Typically, the number of fermions is much smaller than the number of bosons. For example, in the case studied in Ref. \cite{Lewkowicz25}, the number of fermions was set to ten, while the number of bosons was on the order of several thousand.

To summarize (see Ref. \cite{Lewkowicz25}), the Bose-Fermi mixture enters different phases depending on the strength of the attraction between bosons and fermions. When the attraction is weak, the bosonic and fermionic clouds overlap and exhibit Friedel oscillations. With stronger attraction, a dipolar Bose-Fermi droplet forms at a critical value of the interaction strength, $g_{BF}$. This occurs because the interaction between all atoms becomes attractive, analogous to the formation of bright solitons or nondipolar quantum droplets in Bose-Fermi mixtures (see Refs. \cite{Karpiuk04, Rakshit19a, Rakshit19b}). Then, the roton instability develops, and the system enters the supersolid phase. This can be confirmed by checking for the possibility of exciting the out-of-phase Goldstone mode (see movies at Ref. \cite{movies24first}). Similarly, recent studies of various Goldstone modes in the excitation spectrum of elongated dipolar bosonic mixtures revealed a supersolid phase in these systems \cite{Kirkby24}. With even stronger boson-fermion attraction, the fermionic cloud breaks up into separate, non-overlapping peaks, whose number changes from two to the number of fermions. The bosonic background, however, remains -- this marks the self-pinning transition (see also \cite{Keller22}). The system remains in the supersolid phase, but with additional phonon-like modes.

We would like to emphasize that adding fermions to a dipolar Bose-Einstein condensate is essential for achieving the supersolid phase. As shown in Ref. \cite{Andreev17}, a system of ultracold dipolar excitons (pure dipolar bosons) in a quasi-one-dimensional geometry experiences roton instability when polarized perpendicular to the symmetry axis (as in our case), which leads to collapse. That paper demonstrates that this roton instability can be prevented by considering three-body repulsive forces in addition to two-body forces. Subsequently, the transition to the modulated density state was observed. In our case, repulsive forces between fermions due to Pauli exclusion and repulsive dipolar interactions between bosons stop the roton instability. Thus, the Bose-Fermi mixture is stabilized and exhibits an enriched supersolid phase (see Ref. \cite{Lewkowicz25}).

In this paper, we study the possibility of inducing a transition to the supersolid phase through a dynamic change in the boson-fermion attraction strength. Our numerical simulations resemble real experiments in which the bosonic scattering length is tuned using ramps in external magnetic fields lasting a few tens of milliseconds \cite{Tanzi19b,Guo19}.

The paper is organized as follows. In Sec. \ref{Method} we introduce the model of a mixture of dipolar Bose-Einstein condensate and degenerate Fermi gas. Then (Sec. \ref{quench}) we discuss the results of numerical simulations in which we increase the attraction between bosons and fermions in time (i.e. perform a quench), forcing the dynamical transition to the supersolid. Sec. \ref{CrandRb} supports the possibility of the appearance of supersolid phases in systems with atoms of smaller magnetic moment. We conclude in Sec. \ref{conclusion}.

\section{Method}
\label{Method}

We consider an atomic Bose-Fermi mixture at zero temperature and assume its many-body wave function is approximated by a product of Hartree ansatz for bosons and the Slater determinant for fermions

\begin{eqnarray}
&&\Psi ({\bf x}_1,...,{\bf x}_{N_B};{\bf y}_1,...,{\bf y}_{N_F}) =  \nonumber \\  \nonumber \\
&& \phantom{aaaa} \psi_B({\bf x}_1)\, \psi_B({\bf x}_2)...\psi_B({\bf x}_{N_B})    \times    \nonumber \\
&& \nonumber \\
&& \phantom{aaaa} \frac{1}{\sqrt{N_F!}} \left |
\begin{array}{lllll}
\varphi_1({\bf y}_1) & . & . & . & \varphi_1({\bf y}_{N_F}) \\
\phantom{aa}. &  &  &  & \phantom{aa}. \\
\phantom{aa}. &  &  &  & \phantom{aa}. \\
\phantom{aa}. &  &  &  & \phantom{aa}. \\
\varphi_{N_F}({\bf y}_1) & . & . & . & \varphi_{N_F}({\bf y}_{N_F})
\end{array}
\right |  \, .
\label{HartreeSlater}
\end{eqnarray}
All bosons occupy the same single-particle state. The bosonic atoms possess a magnetic dipole moment, which is polarized along a direction perpendicular to the axis of the system's symmetry, as determined by the prolate shape of the trapping potential (see Fig.~\ref{panel}(a)). Fermions are treated individually, with single-particle orbitals assigned to each fermionic atom. Since the fermionic sample is spin-polarized, the only short-range interactions considered are those between bosons, and between bosons and fermions. We further assume that bosons repel each other, while bosonic and fermionic atoms attract. The set of three-dimensional equations describing such a system reads:
\begin{eqnarray}
&&i\hbar\frac{\partial \psi_B({\bf r},t)}{\partial t}  =  \left[-\frac{\hbar^2}{2 m_B}\nabla^2 + V_B({\bf r}) + g_B\, n_B({\bf r},t)  \right. \nonumber  \\
&& \left. + g_{BF}\, n_F({\bf r},t)  + \int V_{DD}({\bf r}-{\bf r^{\prime}})\,n_B({\bf r^{\prime}},t)\,d^3{\bf r^{\prime}} \right] \psi_B({\bf r},t)   \nonumber  \\
\label{bosons3D}
\end{eqnarray}
for bosons and
\begin{eqnarray}
i\hbar\frac{\partial \varphi_j({\bf r},t)}{\partial t} & = & \left[-\frac{\hbar^2}{2 m_F}\nabla^2 + V_F({\bf r}) \right.  \nonumber  \\
&+& \left. g_{BF}\, n_B({\bf r},t) \right] \varphi_j({\bf r},t) 
\label{fermions3D}
\end{eqnarray}
for fermions (here $j$ index runs over the whole set of fermionic atoms). Harmonic trapping potentials $V_{B(F)}({\bf r})=\frac{1}{2} m_{B(F)}\left[\omega^2_{B(F)\perp}(x^2+y^2) + \omega^2_{B(F)\parallel} z^2\right]$ are axially symmetric and
elongated along $z$ direction, coupling constants $g_B=4\pi \hbar^2 a_B/m_B>0$ and $g_{BF}=2\pi \hbar^2 a_{BF}/\mu<0$ are related to the scattering lengths $a_B$ and $a_{BF}$, and $\mu=m_B m_F /(m_B + m_F)$ is the reduced mass of bosonic and fermionic atoms. The long-range dipolar interaction term, assuming magnetic moments are polarized along $x$ direction, is given by
\begin{eqnarray}
V_{DD}({\bf r}) & = & \frac{\mu^2_{dip}}{r^3} \left( 1 - 3\, \frac{x^2}{r^2} \right)   \,,
\end{eqnarray}
where $\mu_{dip}$ is the magnetic moment of bosonic atom. The Eqs. (\ref{bosons3D}) and (\ref{fermions3D}), but without dipolar term, were used to study the formation of Bose-Fermi solitons \cite{Karpiuk04,Karpiuk06} whose existence has been recently confirmed experimentally \cite{Chin19}.

We then reduce the geometry of the system to quasi-one-dimensional one (see Fig.~\ref{panel}(a)). Following the standard approach, the condensate wave function $\psi_B({\bf r},t)$ and fermionic orbitals $\varphi_j({\bf r},t)$ are assumed to be in their ground states in radial directions. After integrating over radial dimensions, Eqs. (\ref{bosons3D}) and (\ref{fermions3D}) become
\begin{eqnarray}
&&i\hbar\frac{\partial \psi_B(z,t)}{\partial t}  =  \left[-\frac{\hbar^2}{2 m_B}\frac{\partial^2}{\partial z^2} + V_B(z) + g_b\, n_B(z,t)  \right. \nonumber  \\
&& \left. + g_{bf}\, n_F(z,t)  + \int V_{dd}({z-z^{\prime}})\,n_B(z^{\prime},t)\,d z^{\prime} \right] \psi_B(z,t)   \nonumber  \\
\label{bosons1D}
\end{eqnarray}
and
\begin{eqnarray}
i\hbar\frac{\partial \varphi_j(z,t)}{\partial t} & = & \left[-\frac{\hbar^2}{2 m_F}\frac{\partial^2}{\partial z^2}+ V_F(z) \right.  \nonumber  \\
&+& \left. g_{bf}\, n_B(z,t) \right] \varphi_j(z,t)  \,.
\label{fermions1D}
\end{eqnarray}
All coupling constants get rescaled as $g_b = g_B / (2\pi L_{\perp}^2)$, $g_{bf} = g_{BF} / (2 \pi L_{\perp}^2))$, and $\mu_d = \mu_{dip} / L_{\perp}$, where $L_{\perp}=\sqrt{\hbar/(m_B \omega_{B\perp})}$ (to simplify further analysis we assume equal radial characteristic length scales for bosons and fermions, i.e., $m_B\, \omega_{B\perp}=m_F\, \omega_{F\perp}$). As a result of reducing procedure the dipolar interaction, $V_{dd}(z)$, itself is changed. It splits into $V_{dd}(z)=V_{dd}^{sr}(z)+V_{dd}^{lr}(z)$, where the attractive short-range part equals $V_{dd}^{sr}(z)=-2/3\, \mu_d^2\, \delta(z)$ and the repulsive long-range part reads
\begin{eqnarray}
 V_{dd}^{lr}(z) = \mu_d^2\, \frac{-2 \sqrt{a}\, |z| + \sqrt{\pi}\, e^{\frac{z^2}{4 a}}\, (z^2 + 2 a)\,
{\rm{erfc}} \large( \frac{|z|}{2 \sqrt{a}} \large) }{8\, a^{3/2}}  \,,   \nonumber  \\
\label{Vdd}
\end{eqnarray}
where $a=L_{\perp}^2 /2$ and ${\rm erfc}$ function is the complementary error function. The Fourier transform of $V_{dd} (z)$, which is used to solve Eq. (\ref{bosons1D}) numerically, is \cite{Pawlowski15}
\begin{equation}
\widetilde{V}_{dd}(k) = \mu_d^2\, f(k^2 a) + \frac{1}{3}\, \mu_d^2   \,.
\label{Vkeff}
\end{equation}
Here, $f(k) = k\, e^{k}\, \mathrm{Ei}(-k)$, where $\mathrm{Ei}$ is the exponential integral function.

\section{Quench induced supersolid phase}
\label{quench}

\begin{figure}[thb]
\includegraphics[width=8.5cm]{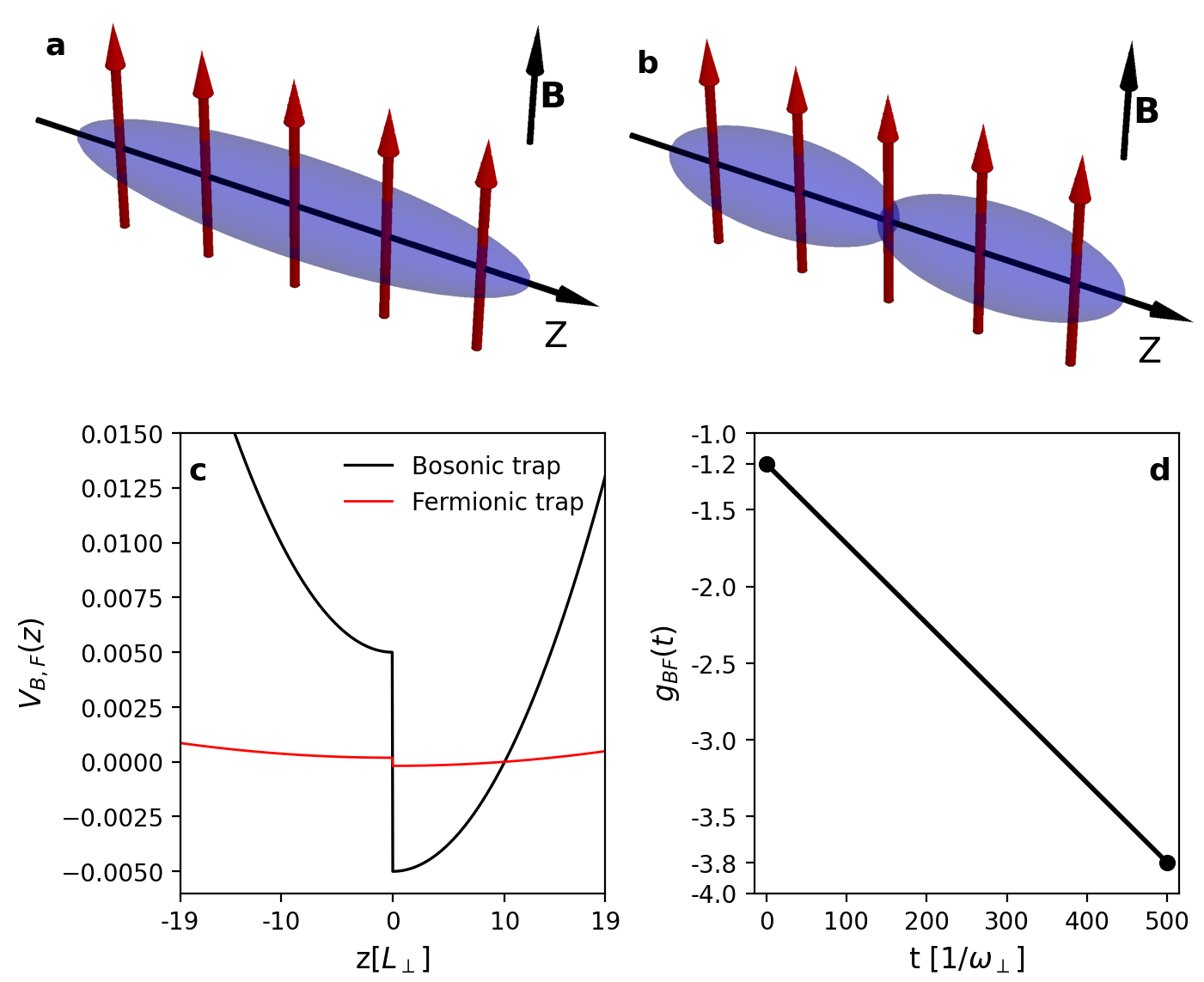}
\caption{(a) Sketch of the physical system under consideration before the quench. (b) A sketch of the system after the quench, when two density peaks indicate the transition to the supersolid phase. (c) Details of the initial trapping disturbance and preparation of the initial state. (d) Details of the quench, including the time dependence of the interspecies attraction.  }  
\label{panel}
\end{figure}

In contrast to Ref. \cite{Lewkowicz25}, we investigate a dynamic method of inducing the supersolid phase in a Bose-Fermi mixture. Following experimental procedures, we vary the strength of the interactions between bosons and fermions over time. Initially, the system is prepared outside the parameter range corresponding to the supersolid phase (as discussed in \cite{Lewkowicz25}), within a trapping potential that is perturbed in a way that breaks the parity symmetry of the system Hamiltonian (see Fig.~\ref{panel}(c) and \cite{Lewkowicz25}). Subsequently, the coupling parameter $g_{BF}$ is varied dynamically to reach the supersolid regime. As expected, in the final state, we observe characteristic density modulations (see Fig.~\ref{panel}(b)). Breaking the parity symmetry in the initial state is intended to trigger excitations characteristic of the supersolid phase. Indeed, these excitations can be observed, but only when the quench time is appropriately chosen. In particular, we can explicitly observe the out-of-phase and in-phase Goldstone modes, as well as the Higgs mode.

All numerical results presented below are intended to model a mixture of fermionic $^{6}$Li and bosonic $^{162}$Dy atoms. However, in Sec. \ref{CrandRb}, we present scaling arguments showing that mixtures containing less magnetic atoms, such as chromium and even rubidium, can also exhibit the supersolid phase. Other system's parameters are: $N_B=3000$, $N_F=10$, and $\mu_d=0.06$ (in units of $(\hbar \omega_{B\perp}\, L_{\perp})^{1/2}$).

We now study quenching that leads to the supersolid phase appearing in different regimes, as identified in Ref. \cite{Lewkowicz25}, we consider two cases: one with larger boson-boson interaction strengths (i.e., before the pinning transition occurs) and one with weaker interactions where additional phonon-like modes emerge. In the former case, with $g_B = 0.01$ (in units of $\hbar \omega_{B\perp} L_{\perp}^3$), the initial state of the Bose-Fermi mixture is prepared as the ground state with $g_{BF} = -1.2$ (in units of $\hbar \omega_{B\perp} L_{\perp}^3$) in the perturbed trapping potential. The quench to the final value of the boson-fermion attraction $g_{BF}=-3.8$ is linear in time and takes $500$ (in units of $1/\omega_{B\perp}$) (see Fig.~\ref{panel}(d)). After quenching, the system enters the supersolid phase (see Fig. 3 in \cite{Lewkowicz25}), which is proved by studying the excitation spectrum. This spectrum is extracted by calculating the Fourier transform of the bosonic density $n_B(z,t)$ in both space and time and integrating out the momentum dependence
\begin{eqnarray}
&&\widetilde{n_B}(\omega) = \int |\widetilde{n_B}(k,\omega)|\, dk  \,,   \label{FT1}
\end{eqnarray}
where
\begin{eqnarray}
\widetilde{n_B}(k,\omega) =  \int \int n_B(z,t)\, e^{i \omega t}\, e^{i k z} dz\, dt  \,,
\label{FT}
\end{eqnarray}
all after the quench is complete. 

\begin{figure}[thb]
\includegraphics[width=7.5cm]{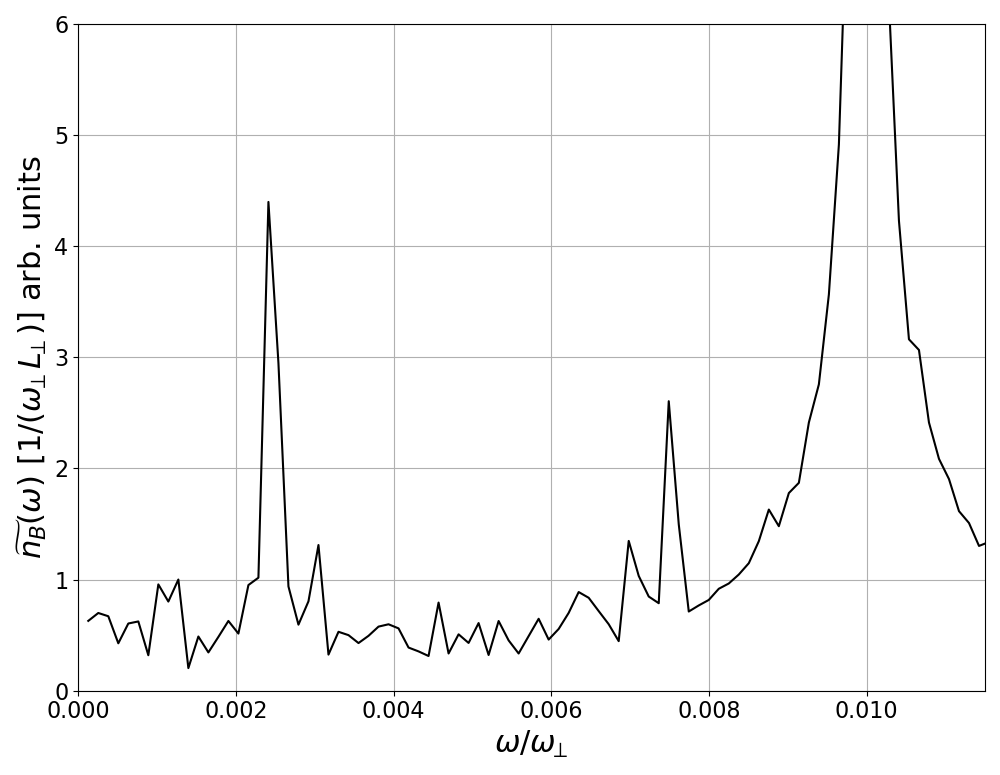}
\caption{Low-energy excitation spectrum of the bosonic component after a quench: $g_{BF}=-1.2 \rightarrow g_{BF}=-3.8$, at constant $g_B=0.01$, within a duration of $500$. Three peaks are clearly visible, representing the in-phase Goldstone mode at $\omega=0.01$ (in units of $\omega_{B\perp}$), the out-of-phase Goldstone mode at $\omega \approx 0.008$, and the Higgs mode at $\omega \approx 0.002$. See movies available at Refs. \cite{movies24,movies25}, showing the dynamics of the Goldstone and Higgs modes.}
\label{spectrum_500_right}
\end{figure}

\begin{figure}[thb]
\includegraphics[width=8.0cm]{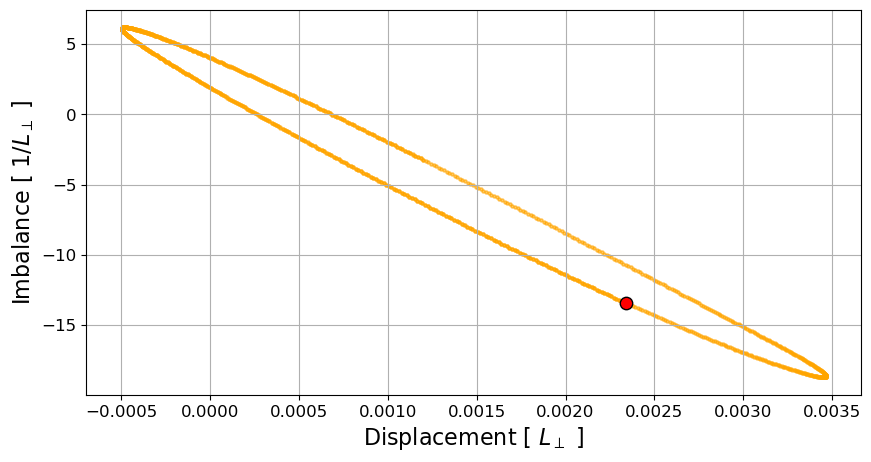}  \\
\includegraphics[width=8.0cm]{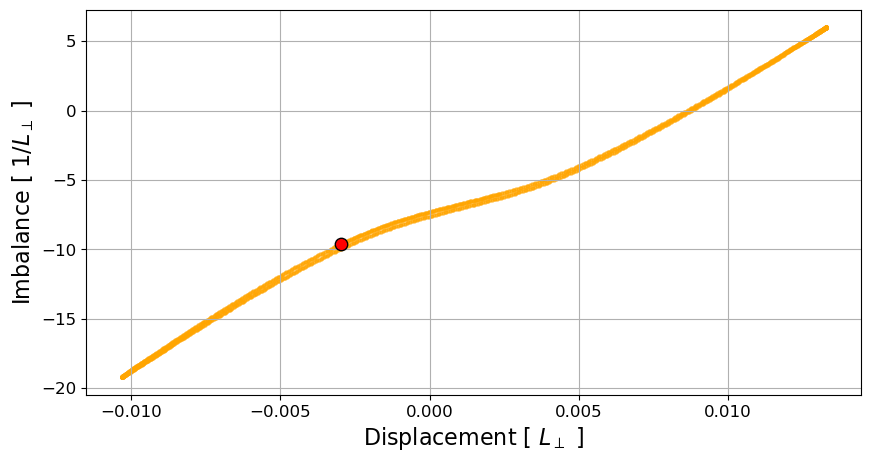}
\caption{Correlation between the imbalance and the displacement for out-of-phase (upper frame) and in-phase (lower frame) Goldstone modes. The modes are excited after quenching: $g_{BF}=-1.2 \rightarrow g_{BF}=-3.8$, with constant $g_B=0.01$, within a duration of $500$. The imbalance for a given mode is defined as the difference in density (obtained from Eq. (\ref{eq5})) maxima between the right and left peaks. Displacement is defined as the average of the left and right peaks center of mass (CM) positions: $(z_{CM}^L+z_{CM}^R)/2$. Here, $z_{CM}^L=\int_{-\infty}^0 z\, n_B^{(\omega_j)}(z,t)\, dz$ and $z_{CM}^R=\int_0^{\infty} z\, n_B^{(\omega_j)}(z,t)\, dz$ and, again, the density $n_B^{(\omega_j)}(z,t)$ is calculated from Eq. (\ref{eq5}). The red points in each frame indicate the mode parameters (imbalance and displacement) immediately after the quench.  }
\label{opgipg_r}
\end{figure}

\begin{figure*}[b!ht]
\includegraphics[width=18.0cm]{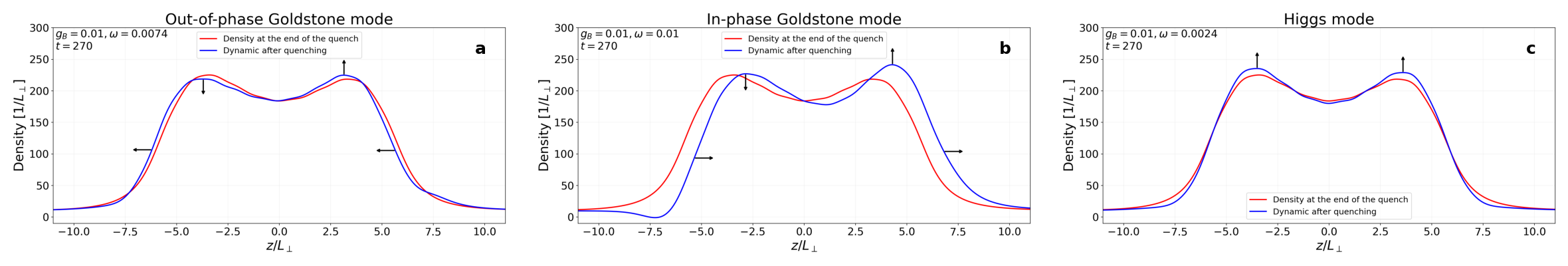}
\caption{(a)-(c) Examples of long-wavelength modes found by the quench. The red curve shows the bosonic density immediately after the quench. The blue curves indicate the densities at specific times and reveal the unique properties of each mode. The horizontal and vertical arrows in frames (a) and (b) show the dynamics of the crystalline and superfluid structures, respectively. When the crystalline structure moves to the left, the superfluid moves to the right, raising the right peak (frame (a)). This is a characteristic of the out-of-phase Goldstone mode. For the in-phase Goldstone mode, both the crystal and the superfluid move in the same direction (see frame (b)). In the case of the Higgs mode, the positions of the density peaks remain essentially motionless while their maxima simultaneously increase and decrease over time. }  
\label{threemodes}
\end{figure*}

For $g_B=0.01$, i.e. in the range of parameters before the self-pinning transition occurs \cite{Lewkowicz25}, the low-energy excitation spectrum is depicted in Fig. \ref{spectrum_500_right}. There are three distinct peaks visible in the spectrum. These are the in-phase, out-of-phase Goldstone and the Higgs modes (with decreasing frequency). The out-of-phase mode is further analyzed in Fig. \ref{opgipg_r} (upper frame). The anticorrelation between the relative heights of the two density maxima (imbalance indicating the flow of the superfluid) and the position of the center of mass of two density peaks (indicating the position of the crystal-like structure) is clearly visible. As the imbalance grows (superfluid flows to the right) the crystal-like structure moves to the left. This anticorrelation behavior is depicted graphically in Fig.~\ref{threemodes} (frame (a)), which sketches the long-wavelength modes excited by the quench.

The out-of-phase nature of the considered mode is also evident in the movie referenced in Refs. \cite{movies24, movies25}. To track the evolution of an excited mode, we calculate the bosonic density (see the Supplementary Material of Ref. \cite{Lewkowicz25} for details)
\begin{eqnarray}
n_B^{(\omega_j)}(z,t) = n_B(z,0) + \overline{n_B} (z,t,\omega_j)   \,,
\label{eq5}
\end{eqnarray}
where
\begin{eqnarray}
\overline{n_B} (z,t,\omega_j) = \left(\int \widetilde{n}_B(k,\omega_j)\, e^{-i k z} dk\right)\,   e^{-i \omega_j t}\,  \,.
\label{eq4a}  
\end{eqnarray}
Once the quench is complete, the out-of-phase mode density represents broken parity symmetry and is indicated by the red point near the lower right vertex of the yellow ellipse in Fig. \ref{opgipg_r}. As the crystal structure moves to the left and the superfluid flows to the right (see \cite{movies24,movies25} and Fig.~\ref{threemodes})), the system moves along the upper branch of the ellipse. When the density peaks are maximally shifted to the left, the flow reverses, and the system moves along the lower branch of the ellipse, exhibiting hysteresis. Conversely, for the in-phase Goldstone mode, the superfluid and the crystal structure move together (see the lower frame in Fig. \ref{opgipg_r}, the movie at \cite{movies24,movies25}, and Fig.~\ref{threemodes})).

\begin{figure}[thb]
\includegraphics[width=7.5cm]{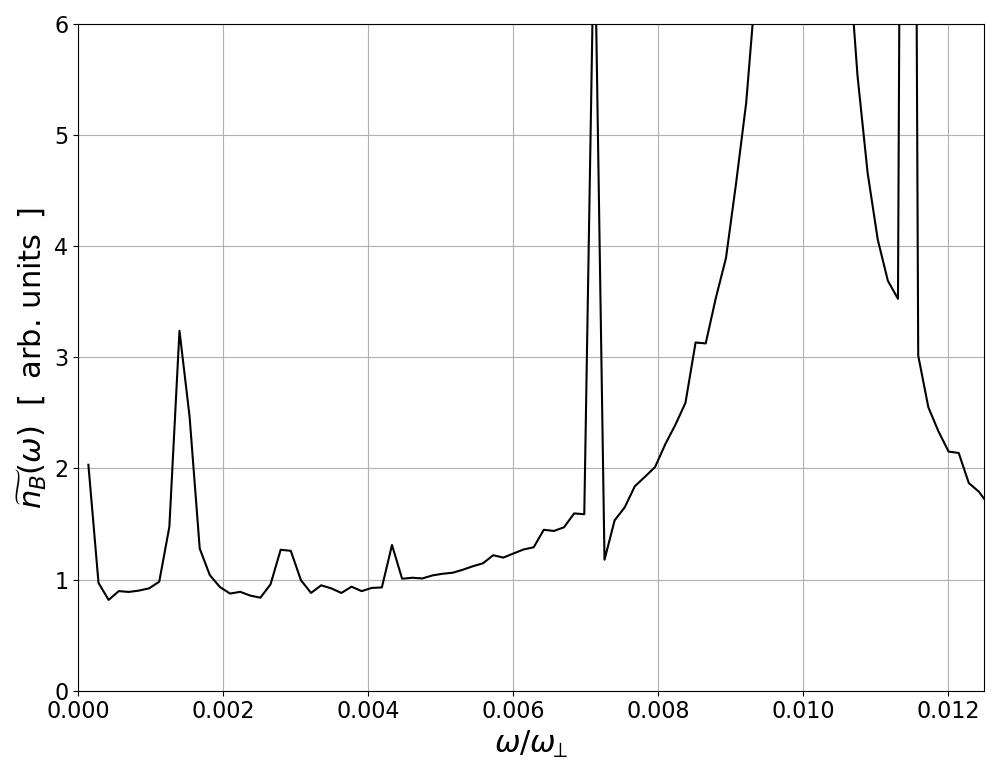}
\caption{Low-energy excitation spectrum of the bosonic component after a quench: $g_{BF}=-3.7 \rightarrow g_{BF}=-3.8$, at constant $g_B=0.0055$, within a duration of $500$. Three peaks are clearly visible, representing the in-phase Goldstone mode at $\omega=0.01$, the out-of-phase Goldstone mode at $\omega \approx 0.007$, and the Higgs mode at $\omega \approx 0.001$. See movies available at Refs. \cite{movies24,movies25}, showing the dynamics of Goldstone modes. An additional peak is visible at approximately $\omega \approx 0.012$, representing the in-phase motion of the two density peaks (see the magenta line in Fig. 3 of Ref. \cite{Lewkowicz25}). This peak is the crystal phonon mode.  }
\label{spectrum_500_left}
\end{figure}

\begin{figure}[thb]
\includegraphics[width=8.0cm]{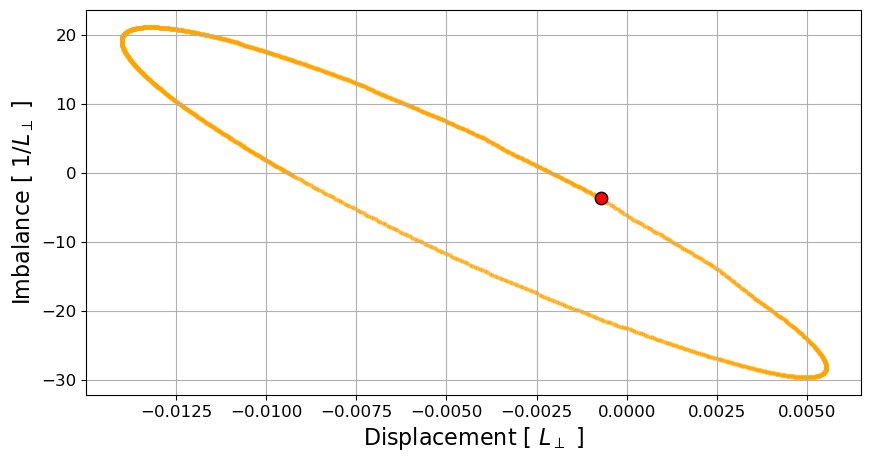}  \\
\includegraphics[width=8.0cm]{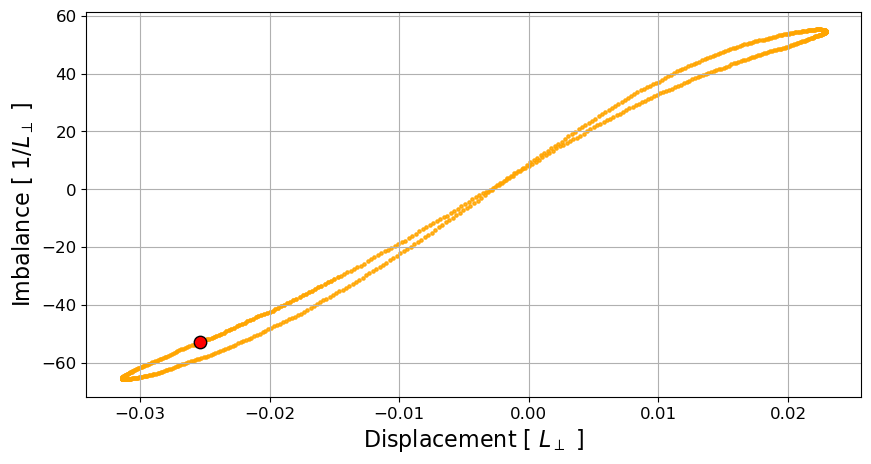}
\caption{Correlation between the imbalance and the displacement for out-of-phase (upper frame) and in-phase (lower frame) Goldstone modes. The modes are excited after quenching: $g_{BF}=-3.7 \rightarrow g_{BF}=-3.8$, with constant $g_B=0.0055$, within a duration of $500$. The red points in each frame indicate the mode parameters (imbalance and displacement) immediately after the quench. }
\label{opgipg_l}
\end{figure}

For the second case, with $g_B = 0.0055$, which belongs to the self-pinning region (see \cite{Lewkowicz25}), the low-energy excitation spectrum is shown in Fig. \ref{spectrum_500_left}. In this case, the initial state of the Bose-Fermi mixture is prepared as the ground state with $g_{BF} = -3.7$, close to the final value of $g_{BF} = -3.8$. The quench is subtle enough to observe the low-energy Goldstone modes, as discussed in the next paragraph in the context of quench duration. Similarly to Fig. \ref{spectrum_500_right}, the distinct peaks represent the in-phase Goldstone mode at $\omega=0.01$, the out-of-phase Goldstone mode at $\omega \approx 0.008$, and the Higgs mode at $\omega \approx 0.002$. The out-of-phase character of the out-of-phase Goldstone mode is clearly visible in Fig. \ref{opgipg_l} (upper frame) and in the movie at \cite{movies24,movies25}.  Additional mode visible at $\omega \approx 0.012$ represents the phonon mode.

In both cases discussed above, the quench duration was equal to 500, as this is the value for which the out-of-phase Goldstone mode is clearly observed in the excitation spectrum. Presence of the out-of-phase Goldstone mode  confirms that at the end of the quench, the system is in the supersolid phase. Interestingly, regardless of the observed density peaks for somewhat different values of the quench rate, we do not find a significant out-of-phase Goldstone mode in the excitation spectrum. This is likely due to the way the system evolves during the variation of $g_{BF}$ over time, particularly in view of the complex excitation spectrum of the Bose-Fermi mixture, which exhibits a number of crossings and anticrossings (see \cite{Lewkowicz25}). A similarly rich excitation spectrum was previously reported for a dipolar dysprosium Bose-Einstein condensate, where crossings or repulsions between Bogoliubov–de Gennes modes were observed as a function of the bosonic scattering length (see \cite{Hertkorn19}). Therefore, depending on the rate of change of the interspecies interaction, the Landau-Zener formula predicts different branching ratios for the occupation of various excited states, so the out-of-phase Goldstone mode can be only barely, if at all, excited if the duration of the quench is not properly adjusted. Our numerical protocol mimics the experimental work, Ref. \cite{Guo19}, in which a quench of a particular duration is performed at the final stage to observe the desired feature in excitations.

\section{Supersolid phase with atoms of smaller magnetic moment}
\label{CrandRb}
The numerical results presented above can be used to model a mixture of fermionic $^{6}$Li and bosonic $^{162}$Dy atoms. The magnetic dipole moment of the bosonic atoms is given by $\mu_{dip}=\tilde{\mu}_d\, \hbar^{5/4}\, (m_B^3\, \omega_{B\perp})^{-1/4}$, where $\tilde{\mu}_d$ ($=0.06$ in the calculations of Sec.~\ref{quench}) is the dimensionless  value of  magnetic moment used in our numerical simulations in quasi-one-dimensional geometry. The value of the magnetic moment equal to $\tilde{\mu}_d=0.1$ (data presented in \cite{Lewkowicz25}) for the trap frequency  $\omega_{B\perp}=2\pi \times 14\,$Hz, corresponds to  $\mu_{dip}=10\,\mu_B$, where $\mu_B$ is the Bohr magneton, i.e. to the magnetic moment of a dysprosium atom. At the same time, the dimensionless values  of $g_B=0.01$ and $g_{BF}=-3.8$ translate to the scattering lengths $a_B=1.7\,$nm and $a_{BF}=-45\,$nm. For $\tilde{\mu}_d=0.06$ to get the value of the magnetic moment of  dysprosium atoms, one has to decrease the radial trapping frequency to $\omega_{B\perp}=2\pi \times 1.8\,$Hz. This leads to larger scattering lengths of $a_B=4.7\,$nm and $a_{BF}=-127\,$nm.

Moreover, since $\mu_{dip} \sim \tilde{\mu}_d\, (m_B^3\, \omega_{B\perp})^{-1/4}$, scaling our results to those of other species is possible. For example, replacing dysprosium with chromium atoms $(m_{B,F} \rightarrow m_{B,F}/3)$ and changing $\tilde{\mu}_d \rightarrow \tilde{\mu}_d/2$ and $\omega_{B\perp} \rightarrow 3^3\,\omega_{B\perp}$, leads to a value of $\mu_{dip}$ that is approximately two times smaller, which is consistent with the magnetic moment of $^{52}$Cr atoms. More precisely, for a radial trapping frequency of $\omega_{B\perp}=2\pi \times 380\,$Hz and a dipolar moment of $\tilde{\mu}_d=0.06$, the formula $\mu_{dip}=\tilde{\mu}_d\, \hbar^{5/4}\, (m_{Cr}^3\, \omega_{B\perp})^{-1/4}$ gives $\mu_{dip}=6\,\mu_B$, which is the magnetic moment of chromium atom. The scattering lengths for interactions between bosons themseves (chromium atoms) and between bosons and fermions ($^{2}$D atoms) are then $a_B=0.6\,$nm and $a_{BF}=-15.2\,$nm, respectively.

Further decreasing the numerical value of $\tilde{\mu}_d$, even systems consisting of non-dipolar gases such as $^{87}$Rb could be considered as exhibiting supersolidity. For a radial trapping frequency of $\omega_{B\perp}=2\pi \times 1000\,$Hz and a value of $\tilde{\mu}_d=0.02$ (unpublished data), the formula $\mu_{dip}=\tilde{\mu}_d\, \hbar^{5/4}\, (m_{Rb}^3\, \omega_{B\perp})^{-1/4}$ gives $\mu_{dip}\approx 1\,\mu_B$, which is consistent with the magnetic moment of a rubidium atom. The scattering lengths for interactions between bosons (rubidium atoms) and between bosons and fermions ($^{3}$He atoms) are then $a_B=0.3\,$nm and $a_{BF}=-7.6\,$nm, respectively.

In general, when planning to work with a particular atomic system, it is important to remember that the scattering lengths, which are determined by the value of the external magnetic field, are $a_B,a_{BF} \sim (m_B\, \omega_{B\perp})^{-1/2}$ (given the dimensionless values of the interaction strengths $g_B$ and $g_{BF}$) and $\mu_{dip}\sim \tilde{\mu}_d\, (m_B^3\, \omega_{B\perp})^{-1/4}$. In order to minimize three-body losses, the scattering lengths $a_B$ and $a_{BF}$ must be kept small, i.e. away from the Feshbach resonance. It is then beneficial to have strong radial trapping and balance the atomic magnetic moment by adjusting the value of $\tilde{\mu}_d$.

\section{Summary}
\label{conclusion}

In summary, we have studied the dynamics of a mixture of dipolar condensate and degenerate Fermi gas in a quasi-one-dimensional geometry, after quenching the interaction between bosons and fermions. For appropriately chosen the final strength of the boson-fermion attraction, the system enters the supersolid phase. The appearance of the supersolid is proved by finding the out-of-phase Goldstone mode in the system excitation spectrum, both for parameters corresponding to the region before and after the self-pinning transition (see Ref. \cite{Lewkowicz25}). We also argue that the supersolid phase can be observed with atoms of smaller magnetic moment.

\acknowledgments  
The authors were supported by the NCN Grant No. 2019/32/Z/ST2/00016 through the project MAQS under QuantERA, which has received funding from the European Union’s Horizon 2020 research and innovation program under grant agreement No. 731473. Part of the results were obtained using computers of the Computer Center of University of Bia{\l}ystok.

\end{document}